\documentclass[twocolumn,twoside,slac_two]{revtex4}
\usepackage{graphicx}
\usepackage{fancyhdr}
\newcommand{\METVEC}{\mbox{$\raisebox{.3ex}{$\not\!$}{\vec E}_T$}}
\def\DZero{D0~}
\newcommand{\MET}{\mbox{$\raisebox{.3ex}{$\not\!$}E_T$}}
\newcommand{\met}{\mbox{$\raisebox{.3ex}{$\not\!$}E_T$}}
\newcommand{\fb}{$fb^{-1}$}
\newcommand{\Et}{\mbox{$E_T$}}
\newcommand{\Pt}{\mbox{$p_T$}}
\newcommand{\et}{\mbox{$E_T$}}
\newcommand{\pt}{\mbox{$p_T$}}

\begin{document}

\title{Observation of Diboson Production in a Semileptonic Decay at CDF}

%

\author{M. Hurwitz, on behalf of the CDF Collaboration}
\affiliation{Enrico Fermi Institute, University of Chicago, Chicago, IL 60637, USA}

\begin{abstract}
We present the first observation of $WW+WZ$ production in the channel
with an identified lepton and two jets in 2.7 fb$^{-1}$ of integrated
luminosity collected with the CDF II detector in $p\bar{p}$ collisions
at $\sqrt{s} = 1.96$ TeV.  The signal is separated from the large
background using matrix element calculations and is observed with a
significance of 5.4$\sigma$.  The $WW+WZ$ production cross section is
measured to be $17.7 \pm 3.1$(stat)$\pm 2.4$(sys)~pb, in good
agreement with standard model predictions.  A complimentary
measurement using a fit to the dijet mass is also presented.
\end{abstract}

\maketitle

\thispagestyle{fancy}


\section{Introduction}

Measurements of heavy vector boson pairs ($WW$, $WZ$, and $ZZ$) are
important tests of the electroweak sector of the Standard Model (SM).
Deviations of the production cross section from SM predictions could
indicate the presence of new particles or
interactions~\cite{hagiwara}.  Furthermore the topology of diboson
events is similar to events where a Higgs boson is produced in
association with a $W$ or a $Z$, so diboson searches and measurements
are useful tests of analysis strategies employed in Higgs searches.

$WW$ and $WZ$ production has been observed at the Tevatron in channels
where both bosons decay leptonically~\cite{diblepCDF}~\cite{diblepD0}.
In semi-leptonic modes, where one boson decays to two quarks, large
backgrounds make the signal extraction more challenging.  CDF recently
reported the first observation of such semi-leptonic decays in a
channel with two jets and large missing transverse
energy~\cite{metjets}.  We present observation of semi-leptonic
diboson decays in a channel with an identified lepton and two jets.
The \DZero collaboration has previously reported evidence of this
signal in 1.07 \fb~\cite{d0lvjj}.

Discrimination of the signal processes from the large backgrounds is
accomplished by use of matrix element probabilities.  The
implementation of this search technique follows that used in the
search for associated Higgs boson production~\cite{WHME}.  A second
complimentary technique fits the invariant mass of the two-jet system
(referred to as the dijet mass or $M_{jj}$).

\section{Experimental Apparatus}

The production is observed in 2.7 fb$^{-1}$ of $p{\bar{p}}$ collision
data with $\sqrt{s}=1.96$~TeV collected by the CDF II detector, which
is described in detail elsewhere~\cite{CDFdet}.  The aspects of the
detector relevant to this analysis are described here briefly.  The
tracking system is composed of silicon microstrip detectors and
open-cell drift chambers inside of a 1.4 T solenoid.  Electromagnetic
lead-scintillator and hadronic iron-scintillator sampling calorimeters
segmented in a projective geometry surround the tracking.  The central
calorimeter covers $|\eta| < 1.1$ while the plug calorimeters extend
the calorimetry into the region $1.1<|\eta|<3.6$.  Outside of the
calorimeter are muon detectors composed of scintillators and drift
chambers.  Cherenkov counters around the beam pipe and in the plug
calorimeter count the inelastic collisions per bunch crossing and
provide the luminosity measurement.

We use a cylindrical coordinate system with its origin in the center
of the detector, where $\theta$ and $\phi$ are the polar and azimuthal
angles, respectively, and pseudorapidity is $\eta= - \rm ln$ $\rm
tan(\theta/2)$. The missing $E_T$ ($\METVEC$) is defined by $\METVEC =
-\sum_iE_T^i{\hat n}_i, i={\rm calorimeter~tower~number}$, where
${\hat n}_i$ is a unit vector perpendicular to the beam axis and
pointing to the $i^{th}$ calorimeter tower.  $\METVEC$ is corrected
for high-energy muons and jet energy corrections.  We define the
missing transverse energy $\MET=|\METVEC|$. The transverse momentum
$p_T$ is defined to be $p\sin\theta$.

\section{Event Selection}

The data samples used in this analysis are collected using trigger
paths requiring a central electron (muon) with \Et (\Pt) $>$~18 GeV as
well as a trigger path requiring two jets and large $\met$.  Offline
we further select events by requiring an electron or muon candidate
with
\Pt$>$20~GeV and exactly two jets with \Et$>25$~GeV.  Jets are
clustered using a fixed-cone algorithm with radius $\Delta R =
\sqrt{(\Delta\eta)^2+(\Delta\phi)^2} = 0.4$ and their energies are
corrected for detector effects.

Several event vetos are imposed to reduce background levels and
achieve good agreement between data and simulation.  We reject events
with any additional jets with $\Et>$12~GeV.  We reduce the backgrounds
due to fully leptonic decays of $t\bar{t}$ and diboson events by
vetoing on additional charged leptons.  We also reject events
containing dilepton pairs consistent with decay from a $Z$ boson,
where the second lepton passes very loose identification cuts.
Cosmic ray and photon conversion vetos are implemented.

We impose additional requirements to reduce the level of background
due to QCD multi-jet events.  These events will enter our sample if a
jet fakes an electron or muon and mismeasurement leads to large
$\met$.  QCD multi-jet events are difficult to model, so we reduce
their contribution as much as possible.  In events with a muon, we
require $\met>$20 GeV and $m_{T}(W)>$10 GeV, where $m_{T}(W)$ is the
transverse mass of the lepton-$\met$ system. Jets are more likely to
fake an electron than a muon, so we impose particularly strict
requirements in events with an electron, requiring $\met>40$ GeV and
$m_{T}(W)>$70 GeV, as well as placing requirements on the $\met$
significance and the angle between the $\met$ and the second jet.

\section{Modeling and Backgrounds}

The dominant background to the diboson signal is $W$+jets production
where the $W$ decays leptonically.  Smaller but non-negligible
backgrounds come from QCD multijet, $Z$+jet, $t\bar{t}$, and single
top production.  The modeling for all processes except QCD multijet
events is provided by event generators and a {\sc geant}-based CDFII
detector simulation.  {\sc Pythia} is used to model $WW$, $WZ$, and
$t\bar{t}$~\cite{pythia} events.  The $W/Z+$jets backgrounds are
modeled using the fixed-order generator {\sc Alpgen} interfaced with
the {\sc pythia} parton showering framework\cite{alpgen}.  {\sc
Madevent} interfaced with {\sc Pythia} is used for the single top
background~\cite{madevent}.  The QCD background is modeled using data
events passing loosened lepton requirements.

The normalization of $WW$, $WZ$, $Z$+jets, $t\bar{t}$, and single top
backgrounds is estimated using predicted or (in the case of $Z$+jets)
measured cross sections and efficiencies derived from simulation.  The
normalization of the QCD background is estimated by fitting the $\MET$
spectrum in data to the sum of all contributing processes, where the
QCD and $W$+jets normalizations float in the fit.  While this fit
provides an initial estimate of the $W$+jets normalization, this
normalization is a free parameter in the final fit used to extract the
diboson cross section.  The expected event yields are shown in
Table~\ref{tab:yields}.  The observed yield agrees well with the total
expected yield since the predicted yields are derived from a fit to
the data.

\begin{table}[ht]
\begin{center}
\caption{\label{tab:yields}Expected event yields for signal and background processes and total observed number of events.}
\begin{tabular}{|l|c|}
\hline
\hline
Process & Event yield \\
\hline
$WW$ signal & 446 $\pm$ 17 \\
$WZ$ signal & 79 $\pm$ 3 \\
$W$+jets & 10175 $\pm$ 152 \\
$Z$+jets & 584 $\pm$ 43 \\
QCD multijet & 283 $\pm$ 113 \\
$t\bar{t}$ & 131 $\pm$ 9 \\
single top & 110 $\pm$ 8 \\
\hline
Observed & 11812 \\
\hline
\hline
\end{tabular}
\end{center}
\end{table}

We expect 525 signal events out of 11812 total events.  The dominant
background is due to $W$+jet events.  The QCD multi-jet background is
small due to the stringent cuts imposed to reduce its size.

\section{Matrix Element Methodology}

A matrix element method is employed to separate signal and background
events.  The probability that an event was produced by a given process
is determined using the leading-order differential cross section for
that process.  For an event with measured quantities $x$, we integrate
the appropriate differential cross section $d\sigma(y)$ over the
partonic quantities $y$ convolved with the parton distribution
functions (PDFs), $f(y_1),f(y_2)$, and over transfer functions
describing detector resolution effects, $W(y,x)$:
\begin{equation}
\label{eqn:evtprob}
P(x)=\frac{1}{\sigma}\int d\sigma(y)dq_1dq_2f(y_1)f(y_2)W(y,x).
\end{equation}
We use the CTEQ5L PDF parameterization~\cite{CTEQ}.  The lepton
momenta and jet angles are assumed to be measured exactly.  $W(x,y)$
is a mapping of measured jet energy to partonic energy derived using
the full detector simulation.  The integration is performed over the
energy of the partons and $p_z^{\nu}$.  The matrix element is
calculated with tree-level diagrams from {\sc
madgraph}~\cite{madgraph}.  Event probability densities are calculated
for the signal processes $WW$ and $WZ$, as well as for the
$Wb\bar{b}$, $Wc\bar{c}$, $Wcj$, $Wjg$, $Wgg$, and single top
($t$-channel and $s$-channel) backgrounds.

The event probabilities are combined into an event probability
discriminant: $EPD=P_{signal}/(P_{signal}+P_{background})$, where
$P_{signal} = P_{WW}+P_{WZ}$ and $P_{background} =
P_{Wbb}+P_{Wcc}+P_{Wcg}+P_{Wgj}+P_{Wgg}+P_{st}$.
We make templates of the EPD for all signal and background processes
and ultimately extract the signal using a fit of the observed EPD
distribution to a sum of the signal and background templates.

\begin{figure}[ht]
\includegraphics[width=0.96\columnwidth]{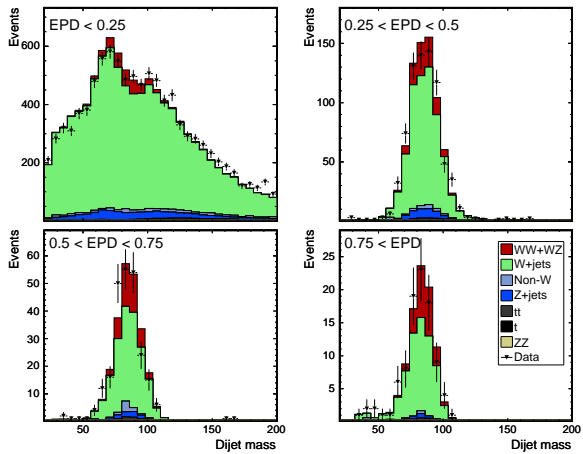}
\caption{$M_{JJ}$ for events with EPD$<$0.25 (top left), 0.25$<$EPD$<$0.5 (top right), 0.5$<$EPD$<$0.75 (bottom left), and EPD$>$0.75 (bottom right).}
\label{fig:EPDbins}
\end{figure}

Figure~\ref{fig:EPDbins} shows the dijet mass in bins of EPD.  Most of
the background events have low EPD, while events with EPD$>$0.25 are
already very signal-like.  The signal-to-background ratio improves
with increasing EPD.

Before comparing the observed EPD to the prediction, we validate the
Monte Carlo modeling of the quantities that enter the matrix element
calculation (the jet and lepton energies and angles).  We divide
events into three regions according to their $M_{jj}$: the signal-rich
region with $55<M_{jj}<120$~GeV/c$^2$ and two control regions with
very little expected signal, $M_{jj}<55$~GeV/c$^2$ and
$M_{jj}>120$GeV/c$^2$.  We compare the observed distributions to the
predicted distributions in these three regions.  In addition to the
matrix element input quantities, we also check the modeling of the
$\met$ and $m_T(W)$ to validate the description of non-$W$ events.
Finally we check $\Delta R_{jj}$, and $\pt_{jj}$ to validate the
modeling of the correlations between the two jets.  All of these
quantities are well-described by the simulation for our event
selection.

A small discrepancy in the modeling of $M_{jj}$ itself can be seen
when combining the two control regions; it is also visible in the
low-EPD region of Fig.~\ref{fig:EPDbins}.  We assign a systematic
uncertainty on the shape of the EPD in the $W+$jets background
associated with this discrepancy by reweighting Monte Carlo events to
make them agree with data.  The weights are derived in the control
regions and extrapolated through the signal region to avoid bias from
the expected signal.  This uncertainty has a negligible effect on the
results, because a large fraction of background events lie in the
first few bins of the EPD distribution.  Small changes in modeling of
those background events do not significantly change the shape of the
EPD.

\section{Systematic Uncertainties}

In addition to the uncertainty due to the $M_{jj}$ mismodeling
described above, we consider several other sources of systematic
uncertainty, taking into account their effect on both the signal
acceptance and the shape of the background and signal templates.  The
uncertainty on the normalization of the backgrounds is considered part
of the statistical uncertainty.  The largest systematic uncertainty is
due to the jet energy scale (JES) uncertainty.  Its effect is
quantified by varying the jet energies in the simulated samples by
$\pm 1 \sigma$.  We assign both an acceptance uncertainty and a shape
uncertainty on the signal templates due to the JES.  Other sources of
systematic uncertainty considered are initial and final state
radiation, parton distribution functions, jet energy resolution, the
factorization and renormalization scale ($Q^2$ scale) used in the {\sc
alpgen} $W/Z$+jets simulations, and the integrated luminosity.

\section{Results}

\begin{figure}[ht]
\includegraphics[scale=0.5,width=1.0\columnwidth]{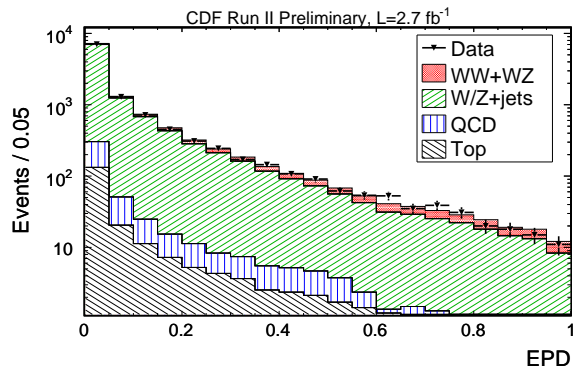}
\caption{Observed EPD distribution superimposed on expected distribution from simulation.}
\label{fig:EPD}
\end{figure}

The observed and predicted EPDs are shown in Figure~\ref{fig:EPD}.  We
use a binned maximum likelihood fit of the observed EPD to a sum of
templates.  Systematic uncertainties are incorporated into the fit as
nuisance parameters.  We perform pseudo-experiments to calculate the
probability ($p$-value) that the background-only discriminant
fluctuates up to the observed result (observed $p$-value) and up to
the median expected signal plus background result (expected
$p$-value).  We observe a $p$-value of $2.1 \times 10^{-7}$,
corresponding to a signal significance of 5.4$\sigma$, where
$5.1\sigma$ was expected.  The observed $WW+WZ$ cross section is
$17.7\pm 3.1$(stat)$\pm 2.4$(sys)~pb.

\section{Search Using Fit to $M_{jj}$}

A complimentary search for a resonance in the invariant mass
distribution has been carried out by fitting the $M_{jj}$ shape to a
sum of signal and background templates.  The background should be
smoothly falling while the signal should exhibit a peak at around
80~GeV from the hadronic decay of a $W$ or $Z$ boson.  This is a
somewhat simpler search technique than using matrix elements as
described above, but is not expected to be as sensitive since it does
not take advantage of the full event kinematics.  A larger data sample
was used for this search, corresponding to 3.9 fb$^{-1}$ of integrated
luminosity.

The event selection for this search is somewhat different than what
has been described for the matrix element analysis.  The central
lepton triggers are used, and offline events are required to contain
one electron candidate with $E_{T} > 20\rm{GeV}$ or one muon candidate
with $p_{T} > 20\rm{GeV/c}$.  We require $\met>25$ GeV and at least
two jets with $\et>$20 GeV.  The non-$W$ background is reduced by
requiring $m_{T}(W)>$30 GeV/$c^2$.  Events are rejected if the two
jets have $\Delta \eta (J1, J2) > 2.5$.  We also reject events whose
hadronic vector boson candidate has $\pt<40$ GeV; this selection
improves the agreement between data and simulation and ensures a
smoothly falling shape in the background dijet invariant mass in the
expected signal region.

\begin{figure}[!htb]
\includegraphics[scale=0.35]{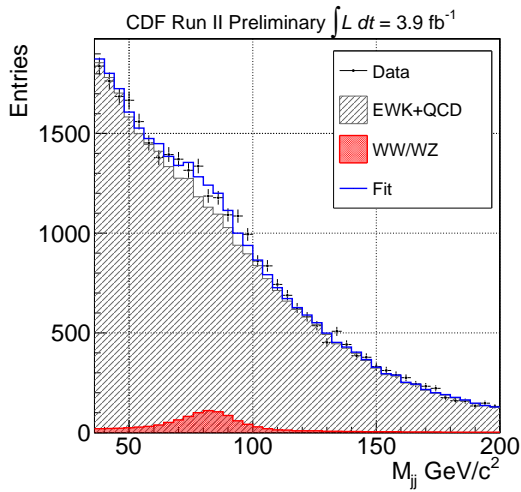}
\caption{\label{fig:projections}Dijet mass distribution in data superimposed on the predicted background and signal distributions.}
\end{figure}

On data, we estimate the signal fraction by performing a $\chi^{2}$
fit to the dijet invariant mass separately for the high $p_{T}$ muon
and electron samples. Three $M_{jj}$ template distributions are used
in the fit.  The first is the combination of $W$+jets, $Z$+jets,
$t\bar{t}$ and single top production (referred to electroweak (EWK)
backgrounds) where the relative normalizations are estimated using
predicted cross sections and efficiencies derived from simulation.
The overall normalization of the EWK contribution is a free parameter
in the signal extraction.  The second template is the multi-jet QCD
background template; its normalization is constrained by the fit to
the $\met$ described earlier.  The third template describes the signal
and its normalization is a free parameter in the fit. In addition, the
global normalization of the sum of our signal and background models is
a free parameter.  Figure~\ref{fig:projections} shows the data
superimposed on the fitted templates after the electron and muon
samples are combined.

The systematic uncertainties in the fit to the $M_{jj}$ distribution
are treated in a similar way as in the matrix element search.  In this
case, however, the largest sources of systematic uncertainty are the
shape uncertainties of the EWK and multi-jet background templates.
The jet energy scale uncertainty is also significant, as is the
overall 6\% uncertainty on the integrated luminosity.

We estimate 428 $\pm$ 177 (stat) $\pm$ 56 (sys) events in the $WW/WZ
\rightarrow e \nu jj$ sample and 650 $\pm$ 149 (stat) $\pm$ 66 (sys)
events in the $WW/WZ \rightarrow \mu \nu jj$ sample, giving cross
sections of 10.3 $\pm$ 4.2 (stat) $\pm$ 1.7 (sys) pb in the electron
sample and 19.5 $\pm$ 4.7 (stat) $\pm$ 2.8(sys) pb in the muon sample.
Combining the two decays, we estimate a total of 1079 $\pm$ 232 (stat)
$\pm$ 86 (sys) $WW/WZ \rightarrow l \nu jj$ events, corresponding to a
significance of 4.6 $\sigma$ and a cross section of $\sigma_{WW/WZ +
X}$ = 14.4 $\pm$ 3.1(stat) $\pm$ 2.2 (sys) pb.  This result agrees
with the cross section measured with the matrix element technique as
well as with NLO prediction.

\section{Conclusions}

In summary, we have observed $WW+WZ$ production in the lepton plus two
jets final state.  We performed two searches: one building a
discriminant using matrix element calculations and the other looking
for a resonance on top of a smoothly falling dijet mass distribution.
The signal was observed with 5.4$\sigma$ significance with the matrix
element search, and $4.6\sigma$ evidence of the signal was found in
the $M_{JJ}$ search.  The $WW+WZ$ cross section was measured to be
$17.7\pm 3.1$(stat)$\pm 2.4$(sys)~pb and $14.4 \pm 3.1$(stat)$\pm
2.2$(sys)~pb with the matrix element and $M_{JJ}$ searches
respectively.  These two results are in good agreement with each other
and with the NLO prediction of $16.1\pm0.9$~pb.


\bigskip
\begin{acknowledgments}
We thank the Fermilab staff and the technical staffs of the
participating institutions for their vital contributions. This work
was supported by the U.S. Department of Energy and National Science
Foundation; the Italian Istituto Nazionale di Fisica Nucleare; the
Ministry of Education, Culture, Sports, Science and Technology of
Japan; the Natural Sciences and Engineering Research Council of
Canada; the Humboldt Foundation, the National Science Council of the
Republic of China; the Swiss National Science Foundation; the
A.P. Sloan Foundation; the Bundesministerium f\"ur Bildung und
Forschung, Germany; the Korean Science and Engineering Foundation and
the Korean Research Foundation; the Science and Technology Facilities
Council and the Royal Society, UK; the Institut National de Physique
Nucleaire et Physique des Particules/CNRS; the Russian Foundation for
Basic Research; the Ministerio de Ciencia e Innovaci\'{o}n, and
Programa Consolider-Ingenio 2010, Spain; the Slovak R\&D Agency; and
the Academy of Finland.
\end{acknowledgments}

\bigskip 

\end{document}